\documentclass[acmsmall, manuscript,screen]{acmart}
\usepackage{multirow}

\AtBeginDocument{%
  }

\setcopyright{acmlicensed}
\copyrightyear{2024}
\acmYear{2024}
\acmDOI{XXXXXXX.XXXXXXX}

\acmConference[Conference acronym 'XX]{Make sure to enter the correct
  conference title from your rights confirmation emai}{June 03--05,
  2018}{Woodstock, NY}
\acmISBN{978-1-4503-XXXX-X/18/06}

\begin{document}

\title{Filling in the Blanks? A Systematic Review and Theoretical Conceptualisation for Measuring WikiData Content Gaps}

\author{Marisa Ripoll}
\affiliation{%
  \institution{School of Computation, Information and Technology, Technical University of Munich}
  \city{Munich}
  \country{Germany}
}
\author{Neal Reeves}
\author{Anelia Kurteva}
\author{Elena Simperl}
\author{Albert Meroño Peñuela}
\affiliation{%
  \institution{Department for Informatics, King's College London}
  \city{London}
  \country{United Kingdom}
}

\author{Klaus Diepold}
\affiliation{
  \institution{School of Computation, Information and Technology, Technical University of Munich}
  \city{Munich}
  \country{Germany}
}

\renewcommand{\shortauthors}{Anonymous}

\begin{abstract}
Wikidata is a collaborative knowledge graph which provides machine-readable structured data for Wikimedia projects including Wikipedia. Managed by a community of volunteers, it has grown to become the most edited Wikimedia project. However, it features a long-tail of items with limited data and a number of systematic gaps within the available content. In this paper, we present the results of a systematic literature review aimed to understand the state of these content gaps within Wikidata. We propose a typology of gaps based on prior research and contribute a theoretical framework intended to conceptualise gaps and support their measurement. We also describe the methods and metrics present used within the literature and classify them according to our framework to identify overlooked gaps that might occur in Wikidata. We then discuss the implications for collaboration and editor activity within Wikidata as well as future research directions. Our results contribute to the understanding of quality, completeness and the impact of systematic biases within Wikidata and knowledge gaps more generally.
\end{abstract}



\received{}

\maketitle

\section{Introduction}

Wikidata is a collaborative knowledge graph launched in 2012 \cite{piscopo2019we} intended as a central knowledge store for Wikimedia projects such as Wikipedia \cite{kaffee2017glimpse}. Consisting of structured data in a database of semantic triples, Wikidata includes almost 14 million entities at the time of writing\footnote{see \url{https://www.wikidata.org/wiki/Special:Statistics}} and has outpaced the English language Wikipedia as the most edited Wikimedia project \cite{piscopo2019we}. In a knowledge graph such as Wikidata, knowledge and information are represented in a graph-based format where nodes represent conceptual (typically real-world) entities and they are connected based on the relationships between those entities \cite{piscopo2019we}. As well as serving as a common source of data for other projects, it also has played a positive role in supporting cross-lingual transfer of information, lowering bot activity and leading to more human collaboration within Wikipedia \cite{vrandevcic2023wikidata}. 

As an open and highly collaborative community, it is dependent on the activity of large numbers of volunteers and bots who regularly make contributions to add information and details about very specific topics \cite{muller2015peer}. The quality of information within Wikidata is reliant on the collaboration of large, heterogeneous groups of volunteers who update information based on their interests and experiences \cite{piscopo2017makes}. This collaboration is generally effective with the quality of information improving steadily over time \cite{shenoy2022study} and the number of contributions made to Wikidata has continued to grow steadily over time \cite{vrandevcic2023wikidata}. 

Yet, there are some significant limitations to this collaboration. Few Wikidata editors contribute across the whole platform and instead prefer to contribute to a limited number of specific areas of interest \cite{muller2015peer, piscopo2017makes}. Contributors are also typically unaware of the use-cases of their contributions due in part to the machine-readable nature of Wikidata and the lack of visibility of this usage \cite{zhang2022working}. This is a problem, because as Wikidata grows, it becomes harder for individual editors and by extension, the wider community to keep track of and monitor deficiencies and gaps within the content \cite{mora2019systematic}. Entities within Wikidata show a long-tail distribution, with many of them having limited number of properties \cite{luggen2021wiki2prop}.

These content gaps have been previously observed and well studied within Wikipedia in areas such as culture \cite{miquel2019wikipedia}, gender \cite{konieczny2018gender}, race \cite{lemieux2023too}, sexuality \cite{ribe2021bridging} and language \cite{miquel2018wikipedia}. There is evidence of content imbalances reflecting these gaps in Wikidata and similar knowledge graphs with consequences extending to other web resources such as search results \cite{radstok2021knowledge}. However, despite these consequences and the attention that Wikipedia gaps have received, there has been limited research to date exploring and quantifying the presence of such gaps within Wikidata.  


Knowledge graphs are very rich data sources as they typically connect and integrate multiple heterogeneous data sources \cite{fensel2020introduction,simsek2022knowledge}. As a result, they have been used as training datasets to create and enhance a range of web-based applications such as question answering, recommendation engines and information retrieval systems such as Google searches \cite{peng2023knowledge,zou2020survey}. Moreover, knowledge graphs are also able to promote explainability\footnote{Models which can be described and explained by human observers in contrast to complex ``black box'' models whose functions are difficult to understand, summarise or even view \cite{xu2019explainable}} AI models and enhance Large Language Models\footnote{Advanced models characterised by their ability to generate and process human language, as well as the large volumes of data involved and high-capacity to learn from inputs \cite{chang2024survey}} \cite{pan2024unifying}. However, imbalances or biases within knowledge graphs, incomplete or outdated information post significant challenges as these may in turn be passed onto the models and technologies developed using such data \cite{tiwari2021recent}. Therefore, a failure to identify and account for these biases and content gaps may have wide-ranging consequences far beyond Wikidata itself. 

In this paper, we present the findings from a systematic literature review of content gaps within Wikidata. Following our methodology presented in section \ref{sec:methodology}, we make three contributions. Firstly (in section \ref{sec:typ}), we present a typology of content gaps for which there is evidence within the literature as well as gaps present within Wikipedia not analysed within Wikidata. Secondly (in section \ref{sec:framework}), we derive a theoretical framework to describe and formulate content gaps and deficiencies within Wikidata and other knowledge graphs. Thirdly (in sections \ref{sec:metrics} and \ref{sec:discussion}), we apply this framework to the results of the literature review to further understand the types of gap present within Wikidata and to identify theoretical gaps and potential research directions for quantifying and addressing these gaps. Overall, we believe that our results contribute to understanding and improving the quality of knowledge and its representation in collaborative knowledge graphs such as Wikidata.



\section{Background and Related Work}

In this section we describe the nature of Wikidata and the structure of the entities and connecting relationships which form the core of the knowledge graph.  

\subsection{Wikidata}
Wikidata\footnote{\url{https://www.wikidata.org/wiki/Wikidata:Main_Page}} is one of the largest open and free collaborative knowledge graphs representing data about more than 100 million concepts as a result of the contribution and collaboration of over 560,000 editors \cite{vrandevcic2023wikidata,vrandevcic2014wikidata}. In contrast to other Wikimedia projects, it provides structured and contextually enhanced data in an easily interpreted format readable by both humans and machines. In Wikidata, items represent real-world things (e.g. topics, people, objects) with labels (i.e. common name for the item), descriptions and aliases (i.e. alternative names). This structured representation supports different agents on the web (e.g. humans, large language models (LLMs) and AI bots) by providing context and meaning in tasks such as information retrieval \cite{geiss2015beyond}, question generation and answering \cite{han2022generating}, collaborative knowledge engineering \cite{piscopo2018models} and language translation \cite{turki2017using}. 

As a highly-collaborative platform, Wikidata relies on contributions from  different users who can have different roles in the leadership and development of the infrastructure (some focusing on defining the underlying schema of Wikidata and others who edit and create content based on it) \cite{zhang2022working}. The involvement of AI bots is another occurring trend in Wikidata content creation and completion \cite{zheng2019roles, erenrichpsychiq}. Due to this, Wikidata's content can often be inconsistent with noticeable uneven coverage of topics \cite{piscopo2018models}. This can potentially result in negative societal impacts such as bias in decision-making due to misrepresentations or even a lack of information (i.e. gaps) in the first place. In contrast to other knowledge graphs, Wikidata does not feature a formally defined ontology\footnote{A contextual framework for what items can be recorded in Wikidata and what relationships they can have to one another. See \cite{piscopo2018models} for further details.} and it is instead constructed on an ad-hoc basis by the collaborative actions of volunteers leading to inconsistencies and variations in its coverage and format \cite{piscopo2018models}.

\subsection{Item Structure}

\begin{figure}
    \centering
    \includegraphics[width=\linewidth]{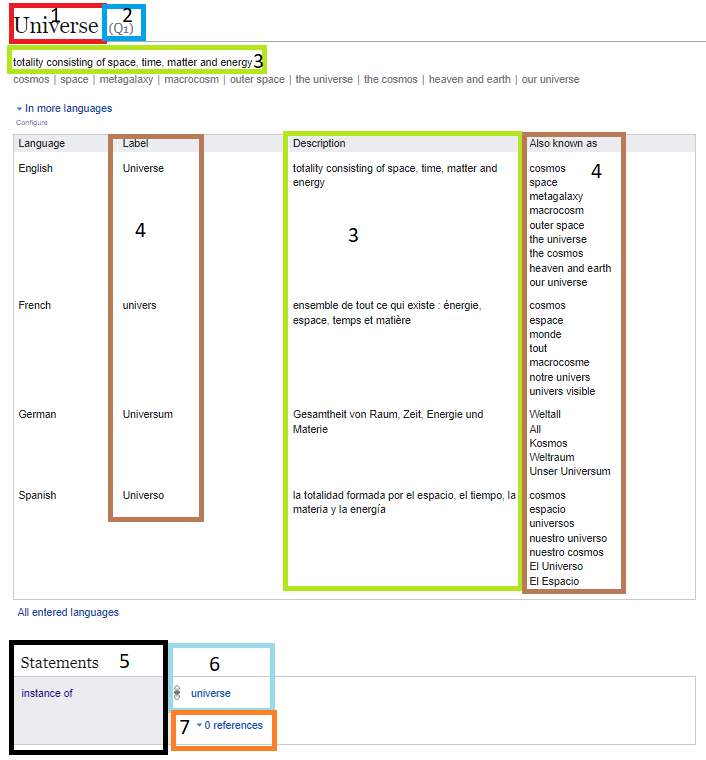}
    \Description[Screenshot of Universe Entity Page]{A screenshot of the entity page for Wikidata entity Q1 -- the Universe. Image shows the entity title, description, multilingual titles and properties, as well as boxes highlighting key sections of the page.} 
    \caption{Wikidata entity page for Q1 -- The Universe.}
    \label{fig:uni}
\end{figure}

An example of the Wikidata entity for the \textit{Universe}\textit{ (Q1)} can be seen in figure \ref{fig:uni}. The item consists of a title (`Universe', labelled 1), entity code (`Q1', labelled 2) and a short description as well as descriptions for other languages (labelled 3). In addition, each language has a main label and a number of alternative labels marked `also known as' shown as number 4 in the figure. The main knowledge-graph content can then be seen under the heading `statements', with the property labelled with number 5 (`instance of'), the associated value with the number 6 (in this case, `universe') and then potentially one or more references which we label 7 in the figure. For ease of reference, we refer to the human-readable, natural text sections of Wikidata listings (labelled 3 and 4) as \textit{descriptive} content while the machine-readable knowledge-graph sections (labelled 5 and 6) we refer to as \textit{structural} content.

\begin{figure}
    \centering
    \includegraphics[width=0.5\linewidth]{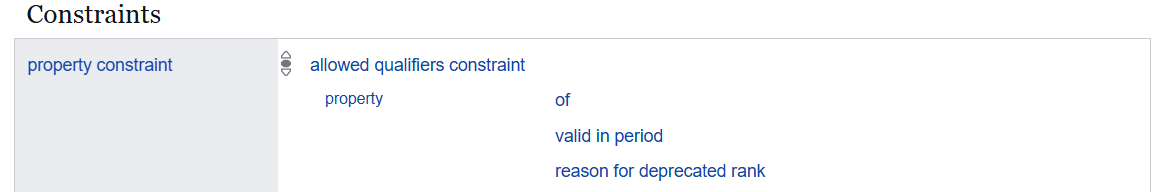}
    \Description[Property constraint screenshot]{Screenshot of a property constraint ``allowed quantifiers constraints'' as seen in WikiData.} 
    \caption{Example of a Wikidata Property Constraint.}
    \label{fig:constraint}
\end{figure}

Properties represent the connections between Wikidata items. For example, the Universe is an instance of a universe but is also part of a multiverse, another entity within Wikidata. Properties have specific constraints which represent machine readable (and enforceable) restrictions which impact the way properties can be applied and used. An example of such constraints for the property \textit{`instance of'} can be seen in figure \ref{fig:constraint}. Properties can also feature qualifiers which provide additional context and detail for relationships. The Wikidata help pages give the example of Emma Watson who has the property cast member within the Harry Potter films, but in turn has the qualifier property of character role for which the entity value is Hermione Granger\footnote{See \url{https://www.wikidata.org/wiki/Help:Qualifiers}}. 

\subsection{Related Work}

Prior work in this space has focused predominantly on the presence of content gaps within Wikipedia. Due to the wide variety of studies within this space, some of which we have already outlined, we do not describe the variety of gaps that have been found. We note, however, that in particular, social biases have been shown to arise in the biographical content associated with articles on human subjects \cite{field2022controlled}. Gaps arise not only in content, but also in the nature of the community with systematic biases and imbalances in the gender \cite{cabrera2018gender}, education \cite{shaw2018pipeline}, age \cite{worku2020exploring} and cultural background \cite{konieczny2020macro} of Wikpedia editors. As well as studying this, we also note relevant literature proposing tools and systems to help address these gaps. A notable example is the work of \citet{miquel2021wikipedia} who developed the Wikidata Diversity Observatory to identify, measure and visualise content gaps within Wikidata to allow editors to address them.  

Past work in Wikidata has largely focused on gaps in the context of quality. \citet{amaral2021assessing} analysed the quality of sources within Wikidata across languages to understand how these influence properties and claims made but of particular interest to our work was their finding that distinct languages feature distinct thematic content. Similarly, \citet{piscopo2019we} performed a literature review to understand how quality is defined within Wikidata touching on multiple themes of relevance to our work including completeness, accuracy and consistency. More specific to systematic content gaps, \citet{weathington2023queer} explored the challenges associated with the accurately detailing queer individuals and themes within Wikidata.

Furthermore, we note relevant work exploring biases in the embeddings found within knowledge graphs. \citet{demartini2019implicit} used paid crowdsourcing experiments to understand the biases that individual contributors might have, but crucially also proposed a method for identifying such biases. \citet{fisher2020debiasing} proposed methods for debiasing knowledge graph embeddings using Wikidata as an example while also identifying sensitive attributes with which gaps and biases may be associated. Similarly, \citet{bourli2020bias} demonstrated a systematic gender bias within Wikidata embeddings and proposed a method to resolve this. Our work builds on these past studies by aiming to catalogue and identify the types of gap previously identified within Wikidata, the methods used to measure these gaps and areas for further analysis and research.

\section{Methodology}\label{sec:methodology}
This section outlines the methodology used for conducting the systematic literature review process. To summarise, we began by identifying terms to look for and carried out a search for these terms across several databases (outlined in section \ref{dbs}). Next, as presented in sections \ref{selection} and \ref{expansion}, we selected, expanded and evaluated the discovered existing literature. Details for each stage of our methodology are provided in the next subsections. The findings were used to derive a typology of content gaps in Wikidata, which we present in section \ref{sec:typ}

\subsection{Database Search}\label{dbs}

One potential barrier to conducting a systematic review of content gaps within Wikidata is the lack of consistent terminology within the wider literature for describing this phenomenon. During a preliminary exploratory analysis, we identified terms such as 'knowledge gap', 'content gap' and 'bias' all used to describe similar issues within knowledge gaps. 

We combined these terms when conducting our search terms and repeated our search across four databases: the ACM digital library\footnote{\url{https://dl.acm.org}}, Sage\footnote{\url{https://us.sagepub.com/en-us/nam/home}}, Scopus\footnote{\url{https://www.scopus.com/home.uri}} and Web of Science\footnote{\url{https://clarivate.com/products/scientific-and-academic-research/research-discovery-and-workflow-solutions/webofscience-platform/}}. Sage, Scopus and Web of Science were selected due to their coverage of a range of disciplines, while the ACM digital library was added due to a lack of computer-science related results in the initial results. While we also trialed searches across other databases, there were significantly diminishing returns in terms of the results which were almost entirely duplicates and we therefore opted not to analyse further databases. A summary of the search terms and fields used can be seen in table \ref{table:databases}.

\begin{table}
    \centering
    \begin{tabular}{lp{20em}ll}
    \hline
        Database & Search Term & Results\\
        \hline
         ACM Digital Library & [Abstract: "wikidata"] AND [[Abstract: "coverage"] OR [Abstract: "bias"] OR [Abstract: "biases"] OR [Abstract: "gaps"] OR [Abstract: "gap"]] & 20\\ \hline
         Sage & "wikidata" AND (("knowledge" OR "content") AND ("gap" OR "gaps")) OR ("bias" OR "coverage") & 2\\ \hline
         Scopus & ("wikidata" AND ( ( "knowledge" OR "content" ) AND ( "gap" OR "gaps" ) ) OR ( "coverage" OR "bias" )) & 146\\ \hline
         Web of Science & ("Wikidata" AND "coverage" OR ("bias" OR "biases") OR ("gaps" OR "gap")) & 69 \\
         \hline
    \end{tabular}
    \caption{Summary of search terms and results used in systematic review.}
    \label{table:databases}
\end{table}

\subsection{Selection Process and Evaluation Criteria}\label{selection}

Following the literature search, results were analysed to identify suitability for analysis. Results were first combined and duplicates removed before the remaining results were analysed through an iterative process. Firstly, results which could be confirmed to not be peer-reviewed (predominantly book chapters and workshop papers) were removed from the sample. Following this, the title and abstract of each result was analysed to remove clearly irrelevant examples. Following this, the full text of the remaining examples was analysed and irrelevant papers were once again removed. The final set of remaining publications then proceeded to the expansion process described below.

Relevance was judged on the basis of three key factors:

\begin{itemize}
    \item Focus on Wikidata -- publications needed to have Wikidata as a predominant or significant focus of the work. We noted many examples which focused on Wikipedia or other platforms and which discussed Wikidata only in passing without making any substantial contribution to understanding the presence of gaps within Wikidata. These were deemed out of scope of the work.
    \item Empirical evaluation -- publications needed to involve empirical evaluation and analysis of gaps within Wikidata. Theoretical works which theorised some form of gap were deemed to be out of scope.
    \item Quantitative analysis -- our goal was to understand the form and scale of gaps within Wikidata. We noted some publications which used qualitative analyses such as interviews with participants which were used to infer gaps such as interviewing female editors to infer a gender gap. However, we found these results to be largely theoretical and insufficient for objectively judging the presence of gaps within Wikidata.
\end{itemize}

While a total of 216 literature results were identified, 66 of these results were found to be duplicates. Among the remaining results, 15 results were not peer-reviewed papers, 1 result was not available in English and 119 results were deemed to be irrelevant. We therefore identified only 19 results directly related to content gaps within Wikidata. This necessitated further data collection to expand our search results.

\subsection{Expansion Process}\label{expansion}

To expand the literature sample, we used forward and back citation tracing with the 19 publications identified through the systematic review process as a basis. For each publication, we first reviewed the full list of references described within the publication, evaluating each reference against the criteria detailed above. We identified a further 26 relevant publications for a total of 45 papers.

\section{Content Gap Typology}
\label{sec:typ}

In this section we lay out a typology of different forms of content gap present in Wikidata as described in the literature. This list covers only those gaps for which we found direct evidence in Wikidata and is therefore unlikely to be exhaustive. Instead, the typology informs our conceptualisation of a content gap, as well as the providing opportunities for further research and analysis which we expand on in section \ref{sec:fut}. 

\subsection{Demographic Gaps}

The first group of gaps can broadly be understood as \textit{demographic} gaps due to arising in the demographic details associated with human subjects enshrined within Wikidata. The presence and quality of subjects has been associated with the gender and the race of that subject.

\subsubsection{Gender}

The largest body of evidence within the sample focused on exploring the presence of a gender gap within Wikidata. Female-dominated occupations tend to feature less diversity \cite{das2023diversity} and Wikidata features significantly more male-dominated occupations than female-dominated \cite{zhang2021quantifying}. In contrast with Wikipedia, Wikidata features a much more diverse range of gender identity properties and there are ongoing debates about how best to represent such characteristics \cite{weathington2023queer}. Due to studies typically considering gender as a male/female binary, it is not possible to ascertain whether such a gap extends to gender identities. 

Nevertheless, the manifestation of this gap is not consistent. There exist both male and female dominated occupations within Wikidata \cite{bourli2020bias} as well as highly negative labels generally albeit not exclusively assigned to either male or female subjects \cite{das2023diversity}. Wikidata has been found to reflect ``real-world gender diversity'' in contrast to Wikipedia where these biases are amplified \cite{centelles2024assessing}. Indeed, the percentage of entities and properties representing male and female subjects and their quality is similar \cite{zhang2021quantifying}. An analysis of the properties associated with MEPs\footnote{Members of European Parliament} found no significant difference in the number or balance of descriptions of family relations associated with gender \cite{hollink2018using}.

\subsubsection{Race}

While white individuals make up 37\% of the human entities within Wikidata, they represent 84\% of scientists and 71\% of the engineers represented within the knowledge graph \cite{shaik2021analyzing}. Moreover, the representations associated with specific races and ethnicities may re-inforce harmful stereotypes such as the high prevalence of Jewish individuals labelled entrepreneurs or economists \cite{fisher2020debiasing}. It is difficult to separate race and ethnicity from culture- and geography-related gaps which we detail below, but it appears from the literature that there are specific concerns and biases related to and embedded within representations of race in Wikidata.



\subsection{Socio-economic Gaps}

One of the few gaps that has been demonstrated to potentially stem from within Wikidata is a socio-economic gap identified through human development indicators \cite{abian2022analysis}. However, this gap is complex and covers a range of factors such as wealth, human development factors such as literacy, but also issues such as geography, culture and ethnicity \cite{abian2022analysis}. It is therefore not possible to conclusively separate this gap from other gaps within the literature, but socio-economic factors are likely to play a role in the content contributed by editors. 

\subsection{Occupation}

Interlinked with the gender gap in Wikidata is an occupation gap, with some occupations more likely to be covered and featured in higher numbers than others \cite{fisher2020debiasing}. Identifying the presence of gaps within Wikidata associated with occupations is complex, however, due in part to the granularity of descriptors and the requirement for inclusion that an individual is notable \cite{klein2016monitoring}. Despite this, while Wikidata may feature a gap associated with ``everyday'' occupations stemming from this requirement for notability, occupations in Wikidata typically reflect those seen in the United States \cite{klein2016monitoring}.

\subsection{Geographic Gaps}

Three types of inter-related gaps emerged around geography and nationality. These included \textit{citizenship}, \textit{culture} and \textit{language}. We detail these gaps as described -- and with the terminology used -- by the authors of the respective publications. However, we note that it is possible and arguably likely that these gaps represent similar phenomena and may even have similar causes. 

\subsubsection{Citizenship}

\citet{abian2022analysis} found a socio-economic gap in the citizenship of individuals represented in Wikidata based on their country of citizenship noting that this was larger than would be suggested by users' information needs. Some forms of gender bias may in fact stem from citizenship. \citet{hollink2018using} found that gender differences in the descriptions of MEPs in Wikidata were typically associated with their country of citizenship and the diversity of its European Parliament and/or Wikipedia content rather than a sign of a gender bias. Similarly, we observed that descriptions of a cultural bias (described below) could not be clearly distinguished from a citizenship bias for human subjects. 

\subsubsection{Cultural}

Evidence for the presence of a cultural gap within Wikidata is somewhat ambiguous. For example, \citet{konieczny2024quantifying} performed an analysis of 90 million Wikidata items alongside 40 million Wikipedia articles to explore the concept of Americanisation, finding that articles related to western cultures show a level of Americanisation while other articles do not. Nevertheless, it is unclear to what extent any Americanisation derives from Wikidata properties as opposed to merely Wikipedia content. Conversely, an analysis of information covering Western and non-Western artists found that Western subjects had typically four times as many statements while Western artworks had nine times as many statements when compared with non-Western subjects \cite{ahmed2023representation}. This suggests that while Wikidata includes content from a variety of cultures, the volume and quality of such content is typically greater when concerning Western cultures.

\subsubsection{Language}

Wikidata features a significant language imbalance with 11 languages responsible for just 50\% of content and ``most languages have close to no coverage'' \cite{kaffee2017glimpse}. This is worse, however, in the entities themselves, with the ontology and properties featuring a much less significant imbalance \cite{kaffee2017glimpse}. Web content as a whole is similarly imbalanced and there is an observed tendency for editors to edit predominantly in their own language \cite{kaffee2018analysis}. Combining and supporting multilingual knowledge transfer is a key research challenge in the domain of knowledge graphs \cite{peng2023knowledge} and it remains unclear how much of this language imbalance is caused by the integration with Wikipedia, which itself features significant language imbalances and multilingual content gaps \cite{mcdonough2017expanding}.

\subsection{Recency Bias}

While the majority of the gaps covered in the literature focus on identity-based gaps, an additional factor was the presence of a recency bias. Wikipedia has been observed to suffer from a recency bias with significantly greater coverage of more recent events \cite{keegan2013hot}. Moreover, ratios of articles on women to men in Wikipedia are low for subjects born prior to the 19th century but then rise steadily and are forecast to reach parity for subjects born in 2034 \cite{konieczny2018gender}. Despite this, only one study described a recency bias within Wikidata itself. \citet{abian2022analysis} observed that there exists a recency bias in Wikidata, but suggested suggest that this factor stems from the information needs of users and/or from Wikipedia, noting that birth dates of subjects covered in other knowledge graphs based on Wikipedia demonstrate a greater recency bias than Wikidata.

\begin{figure}[htbp]
    \centering
    \includegraphics[width=0.7\linewidth]{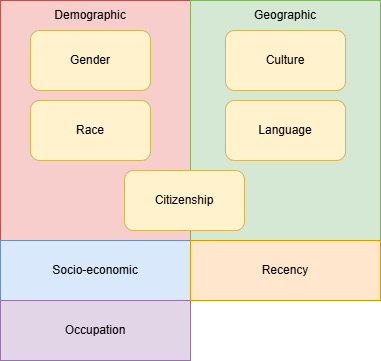}
    \caption{Structural Map of Content Gap Typology present in Literature}
    \Description[Map of Content Gap Typology]{Typology map for content gaps consisting of Demographic (Gender, Race, Citizenship); Geographic (Culture, Language, Citizenship); Socio-economic; Recency; Occupation.)}
    \label{fig:typology}
\end{figure}

\subsection{Manifestation}

Despite the use of the terms \textit{content gap} or\textit{ knowledge \textit{gap}}, we note that these issues can manifest along a diverse spectrum from the complete absence of content to minor missing details within the available content. 

\subsubsection{Presence}

The first and most significant form of gap is the absence of content either due to a lack of representation or a complete failure to represent objects within Wikidata. Less than 22\% of Wikidata items represent women suggesting a systematic absence of content about female individuals \cite{zhang2021quantifying}. Similarly, white, western individuals are over represented compared to the rest of the world \cite{shaik2021analyzing}. 

\subsubsection{Quality}

The quality of content can also involve systemic gaps and biases. Within Wikipedia, for example, articles relevant to specific cultures are typically higher quality and contain more information in languages spoken by those cultures than in other languages \cite{callahan2011cultural}, although perceptions of quality are largely specific to language and culture \cite{jemielniak2017cultural}. More specific to Wikidata, the distribution of properties and information associated with items varies significantly within classes and across the knowledge graph \cite{ramadhana2020user}. Wikidata features ``showcase items'' akin to Wikipedia's featured articles although the number of such items is relatively small and while a quality grading system has been proposed \cite{piscopo2019we}, this does not appear to be integrated directly into Wikidata in the way that Wikipedia's quality grades are. Perhaps because of this, there does not appear to be a single definition of quality within Wikidata and a number of concepts and metrics have been proposed to grade and measure quality \cite{balaraman2018recoin, piscopo2019we,shenoy2022study}.

\subsubsection{Accuracy}

Accuracy can describe both the extent to which Wikidata reflects real-world information (`semantic' accuracy) or the extent to which the content complies with the structural and architectural rules associated with the knowledge graph (`syntactic' accuracy) \cite{farber2018linked}. Typically, however, descriptions of accuracy within the literature focus on semantic accuracy and syntactic accuracy while not actively enforced is monitored and regularly corrected in Wikidata \cite{piscopo2019we}. We therefore focus on semantic accuracy as the likely manifestation of content gaps. Content recorded within Wikidata can be inaccurate either due to human error, lack of activity in particular languages or failure to update outdated (or incorrect) information \cite{conia2023increasing}. 

\subsubsection{Completeness}

A final form that such gaps can take is in the level of detail contained within an entity listing. Wikidata features a diverse array of properties many of which are required only under specific circumstances \cite{balaraman2018recoin} while non-binary properties may be present, accurate and yet incomplete \cite{galarraga2017predicting}. Different languages feature varying levels of detail within their descriptions of Wikidata items \cite{conia2023increasing}, while the number of labels associated with each language is also imbalanced \cite{kaffee2017glimpse}.


\section{Framework}\label{sec:framework}

\begin{figure}
    \centering
    \includegraphics[width=0.4\linewidth]{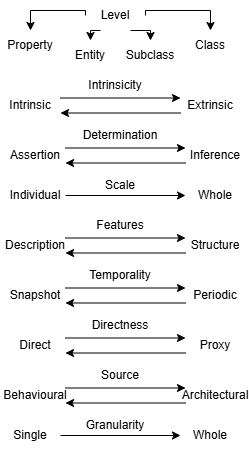}
    \caption{Visual Summary of Theoretical Framework}
    \Description[Summary of framework.]{Visual summary of framework consisting of level (property; entity; subclass; class); intrinsicity (intrinsic/extrinsic); Determination (assertion/inference); scale (individual->whole); features (description/structure); temporality (snapshot/periodic); directness (direct/proxy); source (behavioural/architectural); granularity (single->whole)} 
    \label{fig:theoret}
\end{figure}

In addition to the typology detailed previously, we also identified significant diversity in the way gaps were recognised and measured. In this section, we formulate and describe a theoretical framework aimed to conceptualise and define concept gaps as they arise in Wikidata. Through a set of nine dimensions, we aim to cover the distinct forms such gaps may take, but also the scale and extent of such gaps. A visual summary of this framework can be seen in figure \ref{fig:theoret}.

\subsection{Dimension 1 - Level}

Wikidata content can broadly be conceptualised as a hierarchical model where properties are assigned to items which in turn are assigned (or grouped into) classes, which themselves may be children or `subclasses' of other classes \cite{piscopo2018models}. When identifying, characterising and quantifying gaps it is important to consider the level at which these gaps occur and the level at which they are being measured. Gaps which occur in particular classes may occur in subclasses \cite{berendt2023diversity}, but it is equally possible for a subclass or item to feature distinct gaps from its parent class \cite{ramadizsa2023knowledge}. Despite this, the majority of examples within the sampled studies focused broadly at the property and item level without extending their analysis to classes and subclasses. Understanding the level at which a gap occurs has important implications for the extent and reach of that gap, as well as potential opportunities in indicating the properties, items and classes that may also be impacted by such gaps.

\subsection{Dimension 2 - Intrinsicity}

Not every gap within Wikidata derives from issues within Wikidata itself. For example, while there is a gender gap in Wikidata content, the gap appears to reflect societal biases or biases within the sources used rather than having been introduced or caused by the actions of the Wikidata community \cite{abian2022analysis}. From this we derive our second dimension -- intrinsicity -- designed to capture whether issues are intrisic to Wikidata or not. For the purposes of this dimension, an intrinsic bias or gap represents one that is inherent to Wikidata and/or the actions of editors and is both not reflected in the data source and not an accurate reflection of reality. For example, Wikidata features more entities and properties concerning western artists (and their works) than non-Western artists, but this is not representative of the number or notability of these artists, suggesting that intrinsic factors drive this bias \cite{ahmed2023representation}. Extrinsic knowledge gaps stemming from real-world biases or phenomena -- for example, a lack of female MEPs from some countries \cite{hollink2018using} -- may not truly be `gaps' at all in the conventional sense or at least, may not be possible to address within Wikidata itself. We believe identifying and understanding the intrinsicity of gaps should aid in prioritising those issues which can be addressed within Wikidata as well as potentially identifying external linked sources which may perpetuate these weaknesses.


\subsection{Dimension 3 - Determination}

We note that broadly there are two methods for identifying gaps and errors within Wikidata. The first as covered in much of the literature is to focus on \textit{asserted} knowledge, looking at the relationships and values associated with items to identify whether gaps are present. The second, however, is to infer missing information based on the syntactic structure of the knowledge graph. For example, \citet{shenoy2022study} note that examining constraint violations (i.e., failures to adhere to the syntactic rules) within Wikidata could in theory identify missing triples although in practice the sheer quantity makes this impractical. In fact, some Wikidata items are explicitly recorded as having been inferred from other properties, suggesting that inferred knowledge may play a key role in filling Wikidata gaps \cite{amaral2021assessing}. The use of inference has the advantage of not only identifying gaps, but also resolving them. Nevertheless, we question the extent to which inferring gaps could identify systematic gaps, particularly due to the difficulties of inferring gaps at scale in light of the large number of violations and Wikidata's lack of a formal ontology.

\subsection{Dimension 4 - Scale}

The gaps observed within the literature were typically observable at different levels of scale and can be arranged along a spectrum according to the number of entities required for measurement. For example, it is possible to identify a lack of specifity and detail in textual descriptions of entities by either analysing a single description or comparing different linguistic descriptions of the entity \cite{centelles2024assessing}. Conversely, other gaps can only be identified through the comparison of two or more entities or even through analysing entire subclasses or classes, such as exploring gender disparities \cite{hollink2018using}. As well as governing the measurement of content weaknesses, this factor also indicates the extent to which such gaps occur within the knowledge graph. Analysis of individual entities is insufficient to identify whether gaps are systemic and indicative of biases within Wikidata. The more larger scale the metric, the more likely the metric is to indicate a systematic problem within the knowledge graph and by extension, the more that the metric warrants attention to understand and address the root causes.

\subsection{Dimension 5 - Features}

As demonstrated previously, the fundamental structure of Wikidata entities is a set of triple-based statements which link the entity to other entities or values through particular relationships. However, entities also include a short description of an item in one or more languages which serve to not only help to disambiguate entities, but also to qualify and add context to the data contained within \cite{sardo2022wikidata}.


Evidence from Wikipedia suggests that inequalities can manifest in the language used to describe individuals \cite{wagner2016women, graells2015first}. While we found no evidence of such direct inequalities, Wikidata descriptions and labels feature a number of contentious and potentially offensive terms which may reflect similar inequalities \cite{nesterov2024contentious}. Moreover, the property labels and terminology associated with entities can similarly reflect distinct social biases \cite{fisher2020debiasing}. Focusing solely on structural or textual content, then, has the potential to overlook gaps and biases which may be present in entity listings. 

There are, of course, key differences between the two types of content not only in terms of their nature, but also their applications and significance. Since the triples and structural content are machine readable, they are chiefly used in subsequent applications of the data and therefore more likely to introduce weaknesses and biases than textual content \cite{demartini2019implicit}. 




\subsection{Dimension 6 - Temporality}

An implicit consideration when aiming to quantify and address gaps is the temporary nature with which some biases may be observed. Considering Wikipedia as an example, the death of a human subject may result in a brief period during which an article erroneously describes the individual as alive, but for the average subject editors begin to update articles on the day of death \cite{keegan2015coordination}. Other gaps inherently worsen over time such as a recency bias where current affairs lead to a flurry of edits \cite{keegan2015coordination,bubendorff2021construction} while some gaps appear to be improving over time with the ratio of articles representing women and men increasing steadily over time \cite{klein2016monitoring}. Temporality was inherent in all of the methods used to quantify gaps and biases within Wikidata relying by necessity on the data at the time the calculation was made. However, some analyses explicitly accounted for data over a given time period such as the analysis performed by \citet{abian2022analysis} which accounted for multiple years rather than the data available at a given point.  

\subsection{Dimension 7 - Directness}

Methods for identifying gaps are typically divided into two categories: direct methods which quantify gaps based on specific items and indirect methods which rely on comparisons, theoretical ideals or even proxy measures deemed to approximate a complete or optimal item. This may include comparison with other related entities under the assumption some are more complete than others \cite{ahmeti2017assessing} or may entail the use of proxies for quality such as popularity or lack of alterations \cite{galarraga2017predicting}. While carefully selected comparison entities and properties may be suitable for highlighting content gaps, we note that poorly selected comparisons (particularly for popular entities) risks entrenching and perpetuating content gaps present in such popular content. Conversely, direct methods rarely provide objective insight into the magnitude of a gap and comparative or ideal examples are generally required to truly judge the completeness of an item.

\subsection{Dimension 8 - Source}

Gaps and their measurements also differ strongly in terms of the source or originating factors that lead to the gaps. While some gaps such as gender can be explained by the editing and information seeking behaviours of users \cite{abian2022analysis}, other gaps derive from the underlying architecture or systems defined within Wikidata. For example, while differences in linguistic descriptions partially derive from the behaviours and characteristics of the editing team, they are also influenced by the ontological design and application of Wikidata \cite{kaffee2017glimpse}. Weaknesses and inconsistencies within the ontology impact editors who are constrained by these weaknesses and may subsequently feed into more systematic gaps within content \cite{centelles2024assessing}. Notably, ontological gaps which arise as weaknesses in one source then filter through to other related or linked items \cite{berendt2023diversity}.

Explicitly distinguishing between sources may be more difficult than it first appears, however. In particular, we note the example of schemas, which provide a standardised structure for entities with a common nature within Wikidata. For example, human subjects share a common \textit{human} schema which covers data such as date of birth, citizenship and native language\footnote{See \url{https://www.wikidata.org/wiki/EntitySchema:E10}}. These schemas are defined or supported by specific communities with the knowledge to standardise information related to particular topics \cite{shenoy2022study}. Schemas may be predefined, but they may also arise implicitly through the collaborative actions of the editing community \cite{baroncini2022analysing}. In the case of an implicit schema, then, gaps and weaknesses may arise through adherence to these implicit standards in a form that is not directly codified. 

Furthermore, schemas can be highly specific. Some schemas already encode national or cultural details which may correspond to types of gap such as 'J-Pop Figures' or 'Swedish Soccer Players'\footnote{\url{https://www.wikidata.org/wiki/Wikidata:Database_reports/EntitySchema_directory}}. However, there are a limited number of schemas and they do not effectively reflect every possible permutation of a topic -- the only other schema for soccer players at the time of writing, for example, covers ``Chilean Women's Football players". 

We note that schemas pose both opportunities and challenges in the scope of content gaps. The very presence of schemas has the potential to cause inconsistencies in the quality, scope or even presence of information for subjects from diferent cultures or countries. For example, while there exists a category for Swedish Soccer Players and Chilean Women's Football Players, there are no corresponding categories for other countries leaving individual editors to decide which information should be encoded and how. At the same time, the very presence of these categories facilitates stratified comparisons based on such factors. Were there to be a schema for football/soccer players from every country, it would be possible to easily compare sizes and identify potential gaps within the content or alternatively, it may be possible to encourage consistency in data between demographics, cultures or nations. In the context of our framework, however, they represent something of a middle ground being more architectural and/or structural but potentially only recognisable through the monitoring of behaviours.


\subsection{Dimension 9 - Granularity}

In contrast to the level and scale of gaps, we also observe differences in the granularity at which analyses occurred. These could range from comparing just two properties within an item \cite{conia2023increasing} to analysing the entire knowledge graph \cite{abian2022analysis}. Many of the metrics identified could be applied at a range of granularities, but we note these typically occurred at the level of defined (or assumed) classes such as occupations (e.g., computer scientists and scientists) \cite{centelles2024assessing} or by selecting specific numbers of items \cite{shaik2021analyzing}. While these selections are likely informed by similar considerations as the level and scale dimensions, we nonetheless believe that considering granularity is important to understand the potential repsentativity and limitations of existing and future analyses.


\section{Description of Metrics}\label{sec:metrics}


The metrics used to detect and quantify content gaps in Wikidata lack a standardized framework, making it difficult to compare results across studies and limiting in-depth analysis. Many of the general metrics found in literature can be categorized into either countable or comparative metrics. However, a large amount of previous works have employed case-specific methods to quantify content gaps, which are not applicable to other contexts. A summmary of these metrics can be found in table \ref{tbl:metrics}.

\subsection{Basic}

We refer to basic metrics as metrics used to quantify content gaps based on a countable parameter or an absolute value. These metrics typically used relatively simple calculations (such as simple totals or commonly used calculations such as the mean) to represent and summarise gaps. The use of the term ``basic'' is based on the relative simplicity of the metrics and our belief that these metrics could likely be understood and interpreted by the average Wikidata editor.

\subsubsection{Edits}

Some of the most commonly used methods in literature are countable metrics, with many studies measuring knowledge representation based on various editor-related interactions. Measuring activities such as contributions, revisions, and edits, is a highly versatile option, as it can be effectively applied across a wide range of topics, from social movements \cite{twyman2017black}, to mental health \cite{fraga2018online} and user privacy\cite{rizoiu2016evolution}. That is why, despite the many variations in the terminology used to describe contributions, this metric remains one of the most general and widely adopted methods for identifying content gaps \cite{abian2022analysis}.


We recognise also that analysing the number of edits (or indeed, any contribution) is not in itself sufficient to directly indicate the presence nor scale of a gap. As described by \citet{ahmeti2017assessing}, the absence of edits can in fact suggest a high quality entity listing which does not require updating. Alternatively, Wikidata also features a protection policy which prevents entities from being edited by some types of user which is also likely to impact the number of edits received\footnote{\url{https://www.wikidata.org/wiki/Wikidata:Protection_policy}}.

The number of edits has been used in the literature as a metric to assess classes, properties, and items, reflecting variations in usage according to the level dimension of our framework. As it counts editor activities, this metric is inherently tied to editor behavior, making it an intrinsic measure. It can also help identify and quantify systematic issues at varying scales, allowing for large-scale analysis of Wikidata classes. However, as previously mentioned, this metric alone is insufficient to fully quantify gaps, serving instead as an indirect measure that requires additional variables for comprehensive comparison.

\subsubsection{Diversity of Contributors}

Another proxy measure for the quality of contributions is an analysis of the number of different participants who have contributed to an entity \cite{abian2022analysis}. While this metric is inherently imperfect and it is, of course, possible to have high quality, gap-free content contributed by a small number of users, a lower number of contributors may suggest the presence of gatekeepers \cite{konieczny2024quantifying} and diversity in Wikidata has been previously shown to lead to a higher level of quality \cite{piscopo2019we}.

Like edit counts, contributor diversity can be analyzed at different levels, depending on what contributions are counted (e.g., edits, properties, classes). It is an intrinsic and indirect metric, as it reflects contributor behavior and requires a separate established variable for comparison. Moreover, it can be applied across different scales, using to study both single phenomenons as well as the existence and systematic generation of content gaps.

\subsubsection{Frequency of Contributions}

Calculating the frequency of contributions within a given period or the number of instances (e.g., days) on which edits were received can overcome help to add additional context to more blunt metrics such as edit counts \cite{abian2022analysis}. Nevertheless, we note some potential inconsistencies in the use of such metrics within the literature. For example, the consistency of an entity was seen as a proxy indicator of quality and completeness by \citet{ahmeti2017assessing} while we note that it may equally indicate a stale (and therefore inaccurate or incomplete) entity as described by \cite{conia2023increasing}. Such a metric can vary significantly based on interpretation and context since gaps can occur on a temporal basis. Nonetheless, it remains a useful measure - sharing the same dimensions characteristics as edit counts and contributor diversity. 

\subsubsection{Page views}

Serving largely as a proxy for popularity, page views represent the number of times a page has been visited. This metric has been mostly described in the context of Wikipedia views \cite{abian2022analysis}, but we note that it is equally relevant to Wikidata. Conia et al. \cite{conia2023increasing} compared english vs non-english Wikidata entity names and descriptions making use of Wikipedia pageviews, finding notable differences on the torso and tail entities of the popularity distributions across languages. This work showcases one example of how pageviews can be effective in the analysis of wikidata. More broadly, however, as a proxy for popularity, the use of page views can offer insight into the information needs of the user base which in turn is correlated with some forms of content gap \cite{abian2022analysis}.

In contrast to contributions, page views are largely extrinsic to Wikidata because they reflect user behavior rather than editorial activity. This metric focuses on individual items, measuring how often they are accessed, yet most studies use page views across entire classes to assess content gaps. It is therefore a metric that can be applied to both items and classes. Furthermore, it can be used to assess both one-off and systematic issues. Unlike the previous metrics, page views are more direct, as they provide immediate insight into how often an item is viewed, however they still require interpretation and a comparative variable for comprehensive analysis.

\begin{table}[]
\begin{tabular}{ccllc}
\hline
Type                           & Metric Group                           & \multicolumn{1}{c}{Metric}      & \multicolumn{1}{c}{Associated Gap}           & Source \\ \hline
\multirow{8}{*}{Basic}     & \multirow{3}{*}{Contributions}     & Number of Contributions                     & \multicolumn{1}{c}{\multirow{3}{*}{General}} & \cite{abian2022analysis}      \\
                               &                                        & Diversity of Contributors          & \multicolumn{1}{c}{}                         & \cite{abian2022analysis}      \\
                               &                                        & Frequency of Contributions & \multicolumn{1}{c}{}                         & \cite{abian2022analysis}      \\ \cline{2-5} 
                               & \multirow{2}{*}{Content   Features}    & Number of images per item       & \multicolumn{1}{c}{\multirow{2}{*}{General}} & \cite{zagovora2017weitergeleitet}      \\
                               &                                        & Number of   mentions per item   & \multicolumn{1}{c}{}                         & \cite{zagovora2017weitergeleitet}      \\ \cline{2-5} 
                               & -                                      & Pageviews                       & \multicolumn{1}{c}{General}                  & \cite{abian2022analysis}      \\ \cline{2-5} 
                               & -                                      & Semantic Connections                       & \multicolumn{1}{c}{General}                  & \cite{konieczny2024quantifying}      \\ \cline{2-5} 
                               & -                                      & Byte-Length                     & \multicolumn{1}{c}{General}                  & \cite{ahmed2023representation}      \\ \hline
\multirow{6}{*}{Comparative}   & \multirow{2}{*}{Within   Wikidata}     & Binary                        & \multicolumn{1}{c}{eg: Gender}                              & \cite{abian2022analysis}      \\
                               &                                        & Multiclass                      & \multicolumn{1}{c}{eg: Language}                                      & \cite{kaffee2017glimpse}      \\ \cline{2-5} 
                               & \multirow{2}{*}{Within Wikimedia} &                  Quantifying Linked Content               &                General                              & \cite{luggen2019non}      \\
                               &                                     &      Comparing Content Quantities                           &             e.g., Gender                                 & \cite{klein2016monitoring}      \\ \cline{2-5} 
                               & \multirow{2}{*}{With External Sources}               & Representativity              &  e.g., Gender                                            & \cite{zhang2021quantifying}      \\
                               &                                        & Completeness   & e.g., Gender                                     & \cite{hollink2018using}     \\ \hline
\multirow{5}{*}{Complex} & \multirow{2}{*}{Novel Approaches}        & Embedding Analysis                   &    e.g., Race                                          & \cite{fisher2020debiasing}     \\
                               &                                        & Bias Score                      &      e.g., Gender                                        & \cite{bourli2020bias, das2023diversity}      \\ \cline{2-5} 
                               & \multirow{3}{*}{Defined Metrics}     & Cultural Models                 & Culture                                             & \cite{konieczny2024quantifying}      \\
                               &                                        & Gini   Coefficient              &  Socioeconomic                                            & \cite{ramadizsa2023knowledge}      \\
                               &                                        & Exogenous Measures             &                        e.g., Gender                      & \cite{klein2016monitoring}      \\ \hline
\end{tabular}
\caption{Summary table of metrics identified from the literature review process.}
\label{tbl:metrics}
\end{table}

\subsubsection{Byte-Length}

This metric represents an approximation of the size (and by extension, quality) of an article or entity. The use of bytes facilitates comparisons across languages more easily than other metrics such as character counts as characters in particular languages typically use consistent numbers of bytes\footnote{For example, Latin characters are typically one byte while Hebrew characters use two.} \cite{ahmed2023representation}. Byte length can therefore be an effective method to compare quantities of content between languages. Unlike Wikipedia, Wikidata does not feature multiple instances of each entity based on language, but we note that byte length still represents the number of relations and associated properties that an entity might have and may also be useful for comparing the length of textual descriptions or the number of labels that an entity has.

Byte-length primarily applies to individual items, as it measures how much content is associated with an entity. As a measure of the size of data available for an item, byte-length is an intrinsic metric, tied to how data is stored and represented within Wikidata. It can help identify both one-off content gaps, such as when an individual entity lacks content, and systematic gaps, where entire classes of items have consistently short descriptions.

\subsubsection{Content Attributes: Images and Mentions} 

Other countable parameters that have been used to evaluate items on Wikipedia, and which we believe are also applicable to Wikidata, include the number of images and the number of mentions associated with an item. \citet{zagovora2017weitergeleitet} made an analysis on the gendered presentation of occupations and made use of numbers of images and mentions in Wikipedia articles to measure bias. While Wikidata does not feature mentions in the way Wikipedia does, we believe that links between entities serve a similar purpose and can therefore be measured in the same way. Similarly, Wikidata features links to images within Wikimedia Commons through Property P18, although in contrast to Wikipedia where an article may feature multiple images, the vast majority of Wikidata entities are associated with just a single image \cite{luggen2021wiki2prop}.

Content attributes like images or mentions apply to individual items. These attributes are intrinsic to Wikidata as these content features are directly generated by editors. This metric can be used to detect both isolated gaps (such as a specific item missing an image) and systematic gaps across many entities. However, it remains an indirect metric, as the absence of an image or a mention doesn’t directly reveal a content gap without further contextual analysis.

\subsubsection{Semantic Connections to Wikipedia}

While Wikidata features short descriptions to aid with the disambiguation of different topics, many Wikidata entities are also linked to Wikipedia articles. These semantic connections can be used to aid in identifying concept gaps by examining the long-form content within Wikipedia to better understand biases present in the data. For example, \citet{konieczny2024quantifying} combined an analysis of Wikipedia articles and Wikidata properties to quantify the extent of Americanisation within Wikipedia (and Wikidata). 

Semantic connections apply to items (Wikidata entities) and properties (the entity connection with Wikipedia articles). As for its intrinsicity, while Wikipedias content is external to Wikidata, this metric remains intrinsic to Wikimedia.

\subsection{Comparative}

In contrast to the direct measurements detailed above, we also noted a large number of comparative approaches designed to identify potential gaps through comparison to other resources. These resources derived from Wikidata, other collaborative knowledge sharing platforms such as Wikipedia or exogenous sources of truth.

\subsubsection{Within Wikidata}

The most simple comparisons were performed within and between Wikidata items, classes or descriptions. For descriptive content in particular, the lack of machine-readability and free text nature of the content makes comparisons with other sources (except Wikipedia) difficult. \citet{conia2023increasing} compared the number of errors (particularly typographical errors) and the level of detail between descriptions within Wikidata items to identify language-related gaps and examples where content had been updated in one language but not in another. At the item level, \citet{ahmed2023representation} performed a comparative analysis of the number of properties present within items corresponding to Western and non-Western artists and their works to demonstrate a systematic bias towards Western individuals. At the class level, \citet{abian2022analysis} split classes of entities based on potential gaps (e.g., age, gender) before performing statistical analyses to measure the significance of any potential gap. 

These comparisons within Wikidata used relatively simple values measured with many of the same `basic' metrics as defined above such as edit counts or number of items. However, in contrast to those more basic examples, by performing comparative analyses they were able to offer additional insight into the level and arguably systematic nature of the gap. We also note that comparisons could potentially be made on a temporal basis comparing properties, descriptions or items on a set date with another date or time period, although none of the examples we found chose to do this. Furthermore, comparing at the class level adds an increased sense of scale although other forms of comparison (such as at the description level) will not.

\subsubsection{Within Wikimedia}

In contrast to other knowledge graphs, Wikidata features strong links with Wikipedia, Wikimedia commons and other Wikimedia projects which also support comparative analyses. For example, \citet{luggen2021wiki2prop} were able to analyse images associated with entities purely on the basis that these images were present in Wikimedia Commons and linked from Wikidata using specific properties. This combination of Wikimedia projects can facilitate powerful analyses with \citet{klein2016monitoring} combining the number of Wikidata individuals with the number of biographies present in Wikipedia to understand potential gender gaps present in one or both platforms. Nevertheless, it should be noted that this analysis and the vast majority of the studies found during our sampling process predominantly focused on using Wikidata to understand gaps in Wikipedia rather than vice-versa. Even so, we believe that Wikipedia content could provide vital insight into gaps in Wikidata. Notably, \citet{conia2023increasing} compared descriptions in WiKidata across different languages, but we believe this could also be extended to compare content with Wikipedia to identify where content has been extended in particular platform(s) or language(s) but is not represented in others.

While the use of intra-Wikimedia comparisons provides vital context into the presence and even scale of gaps, we note that it nevertheless may fail to account for intrinsicity. Most notably, if a Wikidata gap is caused by -- or is otherwise a reflection of -- Wikipedia content (e.g., \cite{abian2022analysis}), this within Wikimedia comparison may fail to adequately detect or quantify that gap. There are also difficulties associated with the nature of content particularly within Wikipedia which is more textual and therefore more akin to Wikdata labels and descriptions rather than properties and relationships.

\subsubsection{With External Sources}

The third and final form of comparison noted within the literature was the comparison with external and exogenous sources. Comparison with historical or modern population trends offered insight into the presence of a recency bias and the extent to which Wikidata adequately represents these trends \cite{klein2016monitoring}. Conversely, comparisons with existing knowledge bases were used to assess the completeness and accuracy of information \cite{hollink2018using}. Not all sources were external databases, however, with \citet{konieczny2024quantifying} using a cultural model to identify American topics as a means to quantify Americanisation in Wikipedia and Wikidata.

These external sources provide vital insight into the intrinsicity of gaps and their root causes. While the sources alone do not allow conclusive identification of whether a gap is intrinsic to Wikidata or not, they allow for an understanding of how much Wikidata reflects ``reality'' in a way that is otherwise difficult to assess. However, from the examples we have seen, they are suited only to systematic or larger-scale analyses rather than focusing on granular or small-scale gaps as we assume that the smaller the number of items involved, the more likely to observe a difference from other sources.

\subsection{Complex}

We further identified more complex methods which used either novel formulae and algorithms, artificial intelligence/machine learning, external resources or more complex calculations. Complexity here does not necessarily represent the difficulty of the process required to calculate the metric, but rather represents the predicted familiarity of an average Wikidata editor with such an approach. While we believe Wikidata editors are likely to be familiar with absolute calculations such as total edits and to be able to use or at least interpret such outputs, this is less likely to be true of these more complex methods.

\subsubsection{AI approaches}
Recent advances in Artificial Intelligence have motivated various methods for knowledge graph assessment. A common approach is the use of Knowledge Graph Embeddings (KGE) to measure bias. \cite{bourli2020bias, fisher2019measuring, arduini2020adversarial}
While highly effective at detecting gaps, this method is very case-specific and difficult to generalize. The core issue arises from the embedding generation process, which requires an AI model trained on data tailored to a specific downstream task. Although this approach could be applied across different domains — since it is not inherently tied to a single topic — it still needs separate training for different types of gaps. Developing a more advanced multiclass model could address this limitation; however, no such model currently exists in the literature, presenting a clear opportunity for future research.

Knowledge Graph Embeddings can vary greatly depending on the task. They can operate at the entity and item level when using word embeddings, though even with word embeddings entire subclasses can be analyzed \cite{fisher2019measuring}. Other types of embeddings such as sentence embeddings or document embeddings are also possible. This metric is highly case-specific and can be used to compare large numbers of entities or even classes, indicating a systematic scale. It is intrinsic to the data used to train the embedding models and it is an indirect metric since it requires further analysis by data scientists in order to be used as a metric to analyse content gaps.

\subsubsection{Bias Score}

Originally defined by \citet{fisher2020debiasing}, this metric allows for calculation of biases (e.g., gender bias) associated with particular entities. For example, an occupation may be biased towards a particular gender if many more entities of that gender are linked to the occupation relative to other gender(s). This bias score applies to particular entities but it is used to assess content gaps on the class or subclass level \citet{fisher2020debiasing}. It therefore focuses on a larger scale, with a high granularity aiming to identify systematic issues. It is both intrinsic and direct, since it studies the content within Wikimedia and its function outputs a ratio value that has meaning by itself.

\subsubsection{Exogenous Measures}

\citet{klein2016monitoring} detail four gender equality indicators which differ based in the factors they analyse and the types of inequality they capture. These metrics include the United Nations' Gender Development Index, Social Watches' Gender Equity Index, the Global Gender Gap Index and the Social Institutions and Gender Index. These indices were applied to understand gender inequality in different countries and to represent real-world gender biases which could then be compared with indicators of gender bias in Wikidata (and Wikipedia) to understand whether a gender bias exists relative to wider society.

As external indicators, exogenous measures are naturally extrinsic to Wikidata. These measures are generally considered a complete and reliable source to compare with for the analysis of Wikidata completeness - as such, while these measures tend to be direct and not require further analysis for interpretation, in this case scenario they are indirect - as they only serve Wikidata when used as comparative indicators.

\subsubsection{Cultural models}

The Inglehart-Welzel cultural map, designed by political scientists Ronald Inglehart and Christian Welzel, is a scatter plot and visual representation of cultural patterns around the world. The vertical axis reflects a range from traditional to secular-rational values, while the horizontal axis ranges from survival to self-expression centered values. Another model used to compare cultures is Hofstede's cultural dimensions model, which assesses cultures across 6 dimensions. \citet{konieczny2020macro} applied both models to examine cultural gaps on Wikipedia, finding that contributions tend to be more frequent in countries closer to the self-expression and secular-rational poles of the Inglehart-Welzel map, as well as those scoring higher on Hofstede’s masculinity, uncertainty avoidance, and long-term orientation dimensions.

These extrinsic measures are applied on the item or class level and they uncover systematic gaps by correlating cultural values. Using such models as a metric to detect content gaps is an indirect method, as it makes use of external sociocultural data for analysis and interpretation.

\subsubsection{Gini Coefficient}

The Gini Index or Gini Coefficient is a general tool typically used to measure economic inequality through a single figure ranging between 0 and 1. It represents how equally wealth is distributed within a population. \citet{ramadizsa2023knowledge} applied the Gini Coefficient more broadly to Wikidata by considering wealth as all of the information regarding an entity in the form of relations and objects associated with the entity. By considering entities in a wider class, it becomes possible to evaluate the distribution of information (i.e., properties) associated with the entities within the class.

The Gini coefficient applied to Wikidata is a measure that is implemented at a class or subclass level, since it measures the distribution of information across groups of entities, thus operating on a systematic scale. It is intrinsic, as it reflects inequalities in how information is structured and how content is represented within the platform and it is indirect because it requires interpretation and a comparative benchmarks to detect and quantify content gaps effectively. 




\section{Discussion}\label{sec:discussion}

In this section, we first describe the implications of our findings and particularly the described gaps for collaborative editing behaviours in Wikidata. We also consider the generalisability of our findings and how they might relate to other knowledge graphs and similar systems beyond Wikidata. Furthermore, we present a number of research directions and areas for future work that we believe warrant attention to understand address such systematic gaps not only within Wikidata, but also in other collaborative knowledge sharing systems.

\subsection{Implications for Collaboration}

While content gaps may arise from exogenous factors such as imbalances in external sources or from the underlying architecture and syntactic rules present within Wikidata, our findings also suggest that the individual and collaborative editing behaviours of users contribute to the presence of gaps within Wikidata. A clear example of this is the potential for well-meaning editors who update (and arguably, improve) item descriptions to contribute to gaps in the content due to their inability to update other language descriptions \cite{conia2023increasing, kaffee2017glimpse}. If these weaknesses are to be overcome, then we propose this must be achieved at least partially through new models of collaboration within Wikidata.

To date, editors in Wikidata have largely chosen what to work on based on their own interests and skills \cite{muller2015peer}. This approach is not inherently problematic -- after all, editors are likely to be more motivated by their interests and volunteers should not be expected to work on things they do not wish to do. However, while we found limited evidence surrounding the demographics of Wikidata, the demographics of the overlapping Wikipedia community are somewhat homogeneous. Wikipedia editors tend to be young, well-educated, western males \cite{young2020gender,shaw2018pipeline,graham2015digital}. Whether this applies to Wikidata and whether it influences content gaps and contribution patterns is unclear, but it should be noted that topics of interest to these demographics dominate within Wikidata \cite{konieczny2018gender}.

Whether these demographics and interests directly influence content gaps or not, the quality and quantity of contributions to Wikidata is correlated with the diversity of interests of the contributing community \cite{piscopo2017makes}. Promoting collaboration between diverse users with more diverse interests should ultimately benefit the community as a whole through the generation of more and better quality content. Wikidata already features Wikiprojects with discussions aimed to promote directed collaboration around specific activities \cite{kanke2021knowledge}. However, while Wikiprojects include editing behaviours and facilitate collaboration in Wikipedia, they also require active coordination and leadership on the part of the community if they are to have a significant impact \cite{ren2023did,luyt2018wikipedia}.

\subsection{Generalisability}

This analysis has focused on Wikidata due to its size and scale, its close links with Wikipedia and its potential applications. Nevertheless, we believe our findings are generalisable or otherwise of relevance to knowledge graphs and collaborative knowledge sharing platforms more broadly. The typology described in section \ref{sec:typ} while based on gaps recognised within the literature as occurring in Wikidata (and while likely incomplete) may inform consideration of likely gaps present within other knowledge graphs. Similarly, while the theoretical framework was developed around Wikidata, we nonetheless believe it could inform conceptualisation and detection of gaps across any knowledge graph with a similar structure. Finally, the list of metrics could be applied to a range of similar platforms with many already being applied in e.g., Wikipedia.

\subsection{Research Directions and Future Work}
\label{sec:fut}

We note a number of opportunities for future research and areas of uncertainty within the wider literature. On this basis, we lay out key opportunities and areas where we foresee the research and Wikidata communities would most benefit from further analysis and exploration of the presence and extent of content gaps within Wikidata. 

\subsubsection{Further analysis of content gaps}

In comparison to the content gap typology described in section \ref{sec:typ}, analysis of gaps within Wikipedia and other knowledge graphs has identified a far more diverse array of gap categories. For example, Wikipedia features a content gap concerning LGBTQIA+ people \cite{miquel2021wikipedia} and the Wikidata community has struggled with how to accurately represent gender identity and sexuality to the point of potentially erasing queer identities \cite{weathington2023queer}. Nevertheless, we found no consideration of a sexuality-related gap within the literature while gender-related gaps typically focused on a binary representation which excluded gender identity. An ongoing analysis by \citet{redi2020taxonomy} has identified several other forms of gap likely to occur within Wikimedia projects such as age, sexual orientation and important topics such as health for which we found no Wikidata-specific research. Future work should prioritise investigating the extent and form of content gaps within Wikidata, potentially using prior work carried out on Wikipedia or other knowledge graphs as a starting point.

\subsubsection{Novel and Expanded Analyses}

There exist dimensions within our framework that remain largely theoretical due to an absence of corresponding methods and metrics within the literature. The majority of analyses were performed at the level of individual properties or entities rather than considering classes and few sources identified whether gaps were caused by intrinsic or extrinsic factors. Moreover, while studies focused on asserted knowledge, we believe there are opportunities for inferring gaps based on structural data. Future research in this space should focus on developing novel dimensions, but also combining dimensions such as exploring the temporal nature of observed gaps. 

\subsubsection{Unified Methodologies}

Although we found limited prior research concerning content gaps specific to Wikidata, we nevertheless observed a wide range of methods and metrics for quantifying such gaps. Despite believing further methods are needed, we also note that the diversity in methods poses difficulty for comparing the size and scale of different gaps and biases. Improved commonalities in the methods used or larger scale analyses combining and analysing multiple gaps akin to \cite{abian2022analysis} would be beneficial for identifying the most significant gaps. This in turn would facilitate not only further analysis concerning the source and intrinsicity of such gaps, but also opportunities to resolve or mitigate underlying issues, in turn supporting the commitment of effort and the most effective opportunities for editor time and collaboration.

\subsubsection{Quality Ratings and Gold Standards}

Many of the studies within our sample focused on comparative analyses or calculated discrete values which provided limited insight into gaps without suitable comparisons. However, in the absence of commonly accepted labels for quality and completeness, it was often unclear to which items such comparisons should be made and whether comparative outcomes truly reflected all of the gaps within content. Assessing the quality of Wikidata and linked data knowledge graphs is already an area of significant focus within the literature and we observed a number of studies with robust and accepted methods to evaluate quality across a range of facets (see for example: \cite{piscopo2019we,shenoy2022study,zaveri2016quality}). Nonetheless, while recognised within the research literature, these outcomes are not formally enshrined within Wikikdata. We believe that a set of commonly accepted quality ratings akin to Wikipedia's article rating system and/or the recognition of high quality and complete items, classes or properties could be beneficial for comparative analyses, while adding additional context to direct and less-granular Wikidata analyses. We also believe that benchmarks for formally evaluating and comparing the quality and completeness of Wikidata items could allow for comparisons at scale in a way that is not currently possible. The HELM benchmark for example is able to judge the transparency of Large Language Models but considers concepts such as world knowledge and misinformation \cite{bommasani2023holistic}. Similar benchmarks evaluating the different dimensions and the presence of gaps, biases or inconsistencies within Wikidata could be based on this benchmark and other similar examples.  

\subsection{Editor-Facing Resources}

To support the resolution of gaps and biases within Wikipedia, a range of tools and resources have been developed to help the community find and identify examples. For example, \cite{miquel2021wikipedia} developed the Wikipedia Diversity Observatory consisting of a range of interactive tools and interfaces to visualise gaps to support Wikipedia editors and other stakeholders in ultimately resolving these issues. We found no evidence within our analysis of such tools for Wikidata. Moreover, the studies we found were (understandably) written for an academic audience leaving the results potentially inaccessible for editors and editors potentially unaware of their findings. Many of the metrics described were broadly theoretical or proxy-based (e.g., \cite{galarraga2017predicting}) or relied on concepts such as p-values (e.g., \cite{abian2022analysis}) which editors may not be familiar with. We believe there is a significant divide here between the work of the academic community and the work of Wikidata editors.

If this divide is to be bridged, firstly we believe that the development of tools and/or visual indicators for the community could help them to interface with the outcomes of the ongoing analyses in this space. Secondly, communication and collaboration between Wikidata editors and researchers could provide additional insight into the root causes, challenges and manifestations of gaps, while also providing opportunities to collaborate on resolving such gaps. Dashboards and visualisation tools may be one way to help bridge these gaps, e.g., by replicating the approach used by \citet{miquel2021wikipedia} but with a specific focus on Wikidata.


\subsection{Limitations}

Our analysis was limited by the relatively small number of relevant studies identified during the literature review process. One potential explanation for this was the lack of a consistent vocabulary to discuss such concepts -- for example, while we noted both content and knowledge gap as well as coverage and bias, other sources described similar topics using terms such as completeness, accuracy and consistency. It is therefore possible that our analysis missed relevant studies that used other terminology to discuss such issues. Additionally, our reliance on academic literature for the survey process was inevitably unable to capture the views of the Wikidata community in terms of how they view, conceptualise and monitor gaps. We believe this is an important area for future work.

\section{Conclusion}

In this paper, we presented the results of a systematic literature review around content gaps within Wikidata. We formulated existing research findings into a typology of content gaps observed within Wikidata before describing a theoretical framework intended for conceptualising and identifying content gaps. On the basis of this framework, we then evaluated the metrics and methods used within the wider literature noting dimensions within the framework for which there is limited evidence and where further research and novel methods would be beneficial. While our findings identified a variety of gaps conceptualised in a range of forms, evidence from other knowledge graphs and from Wikipedia shows other potential gaps which have received limited attention in Wikidata to date but which warrant further analysis. We therefore conclude that our understanding of content gaps within Wikidata is currently limited and further work is required to understand the form gaps take, the scale of such gaps and crucially, the root causes which lead to such gaps. Only with a more well-rounded understanding of these issues will it be possible to propose solutions and to collaborate with the Wikidata community to address and resolve these weaknesses.

\bibliographystyle{ACM-Reference-Format}
\bibliography{references}

\end{document}